\documentclass[osajnl,letter,twocolumn,groupedaddress,showpacs]{revtex4}

\usepackage{natbib}
\usepackage{graphicx}
\begin{document}
\title{Two-dimensional atomic lithography by sub-micron focusing of atomic beams}

\author{Will Williams}
\email{wdwilliams@wisc.edu}
\author{M. Saffman}
\affiliation{Department of Physics, 1150 University Avenue, University of Wisconsin-Madison, Madison, WI, 53706}

\date{\today}

\begin{abstract}
We analyze a  method for serial writing of arbitrary two-dimensional patterns using optical focusing of a collimated atomic beam. A spatial light modulator is used in a side illumination geometry to create a localized optical spot with secondary maxima that are well separated from the central peak.  Numerical simulation of a lithography experiment using a magneto-optical trap as a source of cold Cs atoms, collimation and cooling in a magnetic guide, and optical focusing   predicts full width at half maximum pixel sizes of 110 x 110 nm  and writing times of 
about $20~\rm ms$ per pixel.   
\end{abstract}
%\ocis{220.3740,140.3320}
\maketitle

\section{Introduction}
\label{Intro}
Controlling the motion of neutral atoms using light fields has been an important topic in atomic physics for several decades.  Focusing of  atoms from a source onto a planar substrate can be used for lithography where the writing is done by an atomic beam instead of a light field.  This technique is potentially  useful 
 for fabrication of structures with sub-micron resolution. Atomic focusing can be achieved with magnetic or optical fields. In atomic lithography with light fields, an optical profile creates a spatially dependent dipole force that alters the trajectories of neutral atoms. One and two dimensional standing wave light patterns have been used to create periodic atomic patterns\cite{TIMP1992,MCCLEL1993,Schulz2000,Bradle1999,Petra2004}.
Imaging of an atomic beam is also possible using  a magnetic lens as 
was demonstrated by Kaenders et al. \cite{Kaenders1995}, and a wide range of atomic guiding and imaging tasks have been demonstrated using magnetic fields\cite{Hinds1999}.

 While there has been a great deal of work in  atomic lithography (recent reviews can be found in 
Refs. \onlinecite{Oberth2003,Mesche2003,McClel2004}), only spatially periodic or 
quasiperiodic\cite{Jurdik2004,Schulz2000} patterns have been demonstrated. As this limits the range of applications and usefulness of the technique there is considerable interest in devising approaches that will allow spatially complex structures to be created. One approach to creating non-periodic patterns is to use a more complex optical field, as in Refs. \onlinecite{Mutzel2002,Mutzel2003}. An alternative serial writing approach is to focus the atomic beam to a very small spot and then move the spot to draw an arbitrary two-dimensional structure.  Spot motion can be achieved either by scanning the spot over a stationary substrate, or by moving the substrate. An example of an optically scanned atomic beam focused to a size of about  $200~\mu\rm m$ can be found in Ref. \onlinecite{Oberst2003}. To obtain higher resolution 
 tightly focused atomic beams are necessary which can be created by propagation in hollow core fibers\cite{Renn1995} which have the drawback of low atomic flux,  or using  Bessel beams\cite{Bjorkh1978,Balyki1987,Dubets1998,Okamo2001}. A drawback of the Bessel beam approach is the existence of secondary maxima that can lead to atom localization in rings surrounding the central peak.

\begin{figure}[!t]
\begin{center}
\includegraphics[width=5cm]{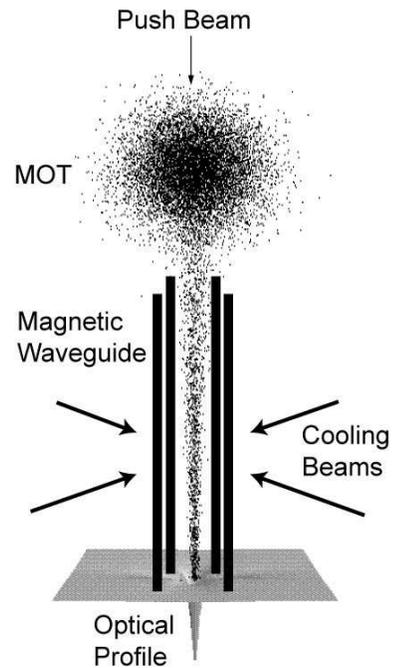}
\caption{\label{fig:scheme}Proposed schematic for focusing neutral atoms to a small isolated dot.}
\label{fig.schematic}
\end{center}
\end{figure}

In this paper we analyze a new approach to focusing an atomic beam to a spot with a characteristic size of about $100~\rm  nm$. The optical profile used for the atomic focusing is created with a spatial light modulator (SLM) that allows the spot to be scanned with no mechanical motion.  By controlling  both the phase and intensity profile of the incident beam we create an optical ``funnel" that 
focuses a high percentage of the atomic flux into a single spot. The proposed approach, as shown in Fig. \ref{fig.schematic}, uses a magneto-optical trap (MOT) as a source of cold atoms.  
A continuous flow of atoms is pushed out of the  MOT\cite{Lu1996,Mandon2000,Mesche2003} and collimated  using a magnetic guide followed by an optical focusing region. The magnetic waveguide creates a micron sized atomic beam, with the final focusing down to a full width at half maximum atomic spot size of $w_{a,\rm FWHM}=110~\rm  nm$ provided by a far detuned  optical profile together with near resonant cooling beams. 
It is then possible to move the optical profile and write a two-dimensional pattern by changing the phases of the laser beams with the SLM. We study the feasibility of this approach using numerical simulations of the atomic trajectories including propagation and cooling  in the magnetic guide and the optical profile. 

In Sec. \ref{opttheory} we summarize the main features of optical focusing of atoms, and describe the creation of a Bessel profile using side illumination with a SLM. The optical funnel for reduction of atomic trapping in secondary maxima is described in Sec. \ref{funnel}. 
The design of a cold atom setup  coupled to a magnetic waveguide and then followed by the optical funnel is described in Sec. \ref{MOT}. Numerical results showing the feasibility of writing a two-dimensional structure are given in Sec. \ref{numerics}.

%%%%%%%%%%%%%%%%%%%%%%%%55
\begin{figure}[t!]
\includegraphics[width=7cm]{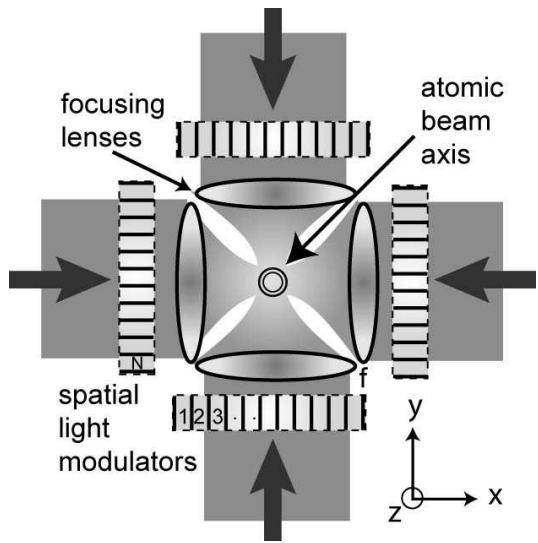}
\caption{\label{fig:slm}Atomic Lithography with a Spatial Light Modulator}
\label{fig.slm}
\end{figure}

%%%%%%%%%%%%%%%%%%%%%%%%%%%%%%%%%%%%%%%%%%%%%%%%%%%%%%%%%%%%%%%%
\section{Optical potential}
\label{opttheory}
\indent The goal of optically mediated  atomic lithography is to control the trajectories of atoms by means of light fields.  A collimated  atomic beam is passed through a region of spatially varying optical intensity that modifies the atomic trajectories such that a desired  atomic pattern is deposited on a substrate. 
  The conservative optical potential for a two-level atom including the effects of saturation is\cite{McClel1995}
\begin{equation}
\label{eqpotential}
U=\frac{\hbar \Delta}{2}\ln\left[1+ \frac{I}{I_s}\frac{1}{(1+4\Delta^2/\gamma^2)}\right],
\end{equation}
where $I_s$ is the saturation intensity, $\gamma$ is the natural linewidth, $\Delta=\omega-\omega_a$ is the detuning from resonance, $\omega$ is the optical frequency, and $\omega_a$ is the atomic transition frequency.
We write the intensity as  $I=I(x,y)g(z)$ where $I(x,y)$ gives the dependence in the $x,y$ plane and  $g(z)$ is an envelope function 
which describes the intensity profile along the $z$ axis, which we will take to be the propagation direction of the atomic beam. Atoms propagating through a region of  spatially varying intensity experience a dipole force ${\bf F}=-\nabla U$ which alters their trajectories, and can be used to focus the atoms into a desired pattern.  When $\Delta<0$ (red detuning) we get an attractive potential that concentrates the atoms where the intensity is highest, while for $\Delta>0$ (blue detuning) the potential is repulsive. For potentials of interest we calculate the atomic trajectories numerically using the classical equations of motion for the atomic center of mass.  It is also assumed that the atoms do not collide and only interact with the given potential.  Therefore, each atom trajectory can be treated individually and a large number of single atom trajectories resulting from a statistical distribution of initial conditions can be  combined  to determine an output  distribution.

%%%%%%%%%%%%%%%%%%%%%%%%%%%%%%%%%%%%%%%%%%%%%%%%%%%%%%%
\begin{figure}[!t]
\includegraphics[width=7cm]{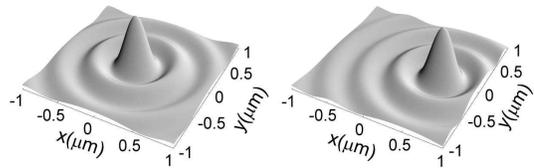}
\caption{\label{fig:phase}Example of translating a $J_0$ Bessel profile composed of 32 laser beams by adding appropriate phase shifts to the interfering plane waves. }
\end{figure}

The  optical potential can be  constructed by combining several laser  beams.  When only a few beams are used the potential has a periodic structure, e.g. a one-dimensional standing wave created by two counter propagating beams, or a checkerboard pattern created by four beams. 
To focus all of the atoms to a single spot, the periodic structure must be removed.  This can be done by adding more laser fields.  Consider a  two-dimensional field formed using $N$ laser beams all propagating in the same plane and arranged to cross at a common point. The shape of the resulting intensity  profile is determined by the angles between the beams  as well as the magnitude and phase of the fields.  The simplest possibility is to cross laser beams, which all have equal electric field phases and magnitudes, with equal angular spacing.  As the number of beams goes to infinity the intensity  profile tends to $J_0^2(k\sqrt{x^2+y^2}),$   the square of the zeroth order Bessel function which has a ring   structure whose scale is dependent only on the wavelength of the light through $k=2\pi/\lambda.$  An axicon can be used to create a Bessel beam for this 
purpose\cite{Dubets1998}, and it is possible to create higher order Bessel profiles  such as a  $J_1$ profile  as proposed by Okamoto et al \cite{Okamo2001}, by altering the phase profile of the beam.

%%%%%%%%%%%%%%%%%%%%%%%%%%%%
\begin{figure}[!t]
\includegraphics[width=8cm]{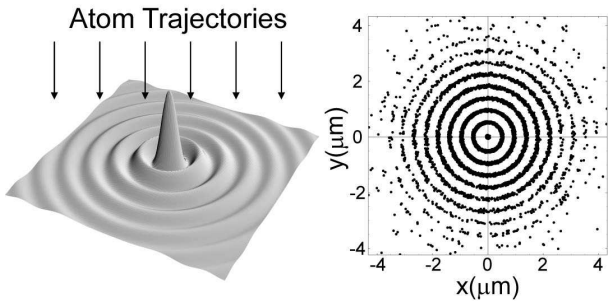}
\caption{\label{fig:bessel}Atom focusing in two dimensions using a standing wave $I\sim J_0^2(k \sqrt{x^2+y^2})$ light mask.   On the right is a numerical example of atom focusing in the light mask. 
The atomic beam had an axial velocity of $14~\rm m/s$, a transverse distribution with  $w_{a,\rm FWHM}=1.6~\mu\rm m$, and a transverse temperature of $T_a=22.5~\mu\rm K.$ The peak potential depth is $U/k_B=21~\rm  mK$ and the optical mask is created with light of $\lambda=.852~\mu\rm m.$ }
\end{figure}
%%%%%%%%%%%%%%%%%%%%%%%%%%

An alternative approach to creating a Bessel beam, as well as higher order beams, is to use a SLM.  There has been substantial recent interest in using SLM technology in atom optics \cite{McGloi2003} as well as an experimental demonstration of manipulation of atoms in microscopic optical 
traps\cite{Bergam2004}. 
Superpositions of Bessel functions may also be useful for addressing individual atoms in optical 
lattices\cite{saffma2004}. 
Referring to the geometry of  Fig. \ref{fig.slm},  when the number of laser beams $N>32$ the pattern is periodic  on scales much longer than the size of the central Bessel peak, so we obtain a well isolated Bessel profile.  
After the optical profile is constructed, serial writing of a pattern can be accomplished by translating the central spot where atoms are focused.   The proposed method to accomplish this is to change the phase of each of the individual beams to construct the profile at a new point on the plane.  Each beam with index $j=(1,N)$  can be represented as an electric field $E_j=E_0 \thinspace \exp\{i k[ \cos(\theta_j)\thinspace x + \thinspace \sin(\theta_j)\thinspace y]-i\,\phi_j\}$ where $E_0$ is the amplitude, $k$ is the wavenumber, $\phi_j$ is a phase, and $\theta_j$ is the angle of the  beam propagation direction with respect to the $x$ axis. We assume all beams are polarized along $\hat z$ so we can neglect vectorial effects.   To create a Bessel profile centered at $x=y=0$ we choose all $\phi_j=0.$ 
To move the  profile to be centered at coordinates $(x_0,y_0)$ we put
\begin{equation}
\label{phase}
\phi_j=k[\cos(\theta_j) x_0+  \sin(\theta_j) y_0].
\end{equation}
This enables translation of the pattern as shown in  Fig.  \ref{fig:phase}. A spatial light modulator can readily be used to create these phase shifts in a side illumination geometry which is compatible with deposition of an atomic beam, as shown in Fig.  \ref{fig.slm}.

 The results of a numerical simulation of atom focusing using a $J_0^2$ Bessel profile can be seen in Fig. \ref{fig:bessel}. In order to reduce heating that occurs when the atoms enter the focusing region two-dimensional near resonant molasses beams  with the same axial profile  as the focusing potential are included.  Details
of the numerical method, including parameters of the  cooling beams, 
are given in  Sec. \ref{numerics}.  A serious problem with this approach is that the atoms are not focused to a single spot.  It is  difficult to obtain a large atomic flux in a beam that is narrow enough to prevent focusing into the surrounding ring structure.   To correct this, the rings of the Bessel function need to be removed.  One possibility is to superimpose a red detuned $J_0$ profile with a blue detuned and repulsive $J_1$ (or higher order) profile. 
The wavelengths and amplitudes of the two Bessel beams can be chosen to suppress the ring structure.  While the first ring can be suppressed, higher order rings are still present and the blue detuned Bessel profile needs to have a very large detuning. This is because the first maximum of higher order Bessel profiles is not at the same radius as the first ring of the zeroth order Bessel.  For example, the third order Bessel has its first maximum at $0.668 \lambda$ while the first ring  of the zeroth order Bessel is at $0.61 \lambda$.  Therefore to get the first maximum to line up with the first ring, we must change the wavelength to $776~\rm  nm$, as compared to the $852~\rm nm$ used for  $J_0$ focusing with Cs atoms.  Since the potential is proportional to $1/\Delta$ for large detuning, the power required to obtain the correct well depth to cancel the first order ring is extremely large.

\section{Optical Funnel}
\label{funnel}

%%%%%%%%%%%%%%%%%%%%%%%%%%%%%%%%%%%%%%%%%%%%%%5
\begin{figure}[!t]
\includegraphics[width=8cm]{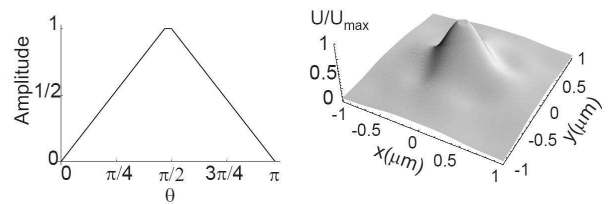}
\caption{\label{fig:funnel}The Optical Funnel profile has wave amplitudes that subtend an angular range of $\pi$ (left), resulting in a localized potential well without secondary rings (right).}
\end{figure}

\begin{figure}[!t]
\includegraphics[width=8cm]{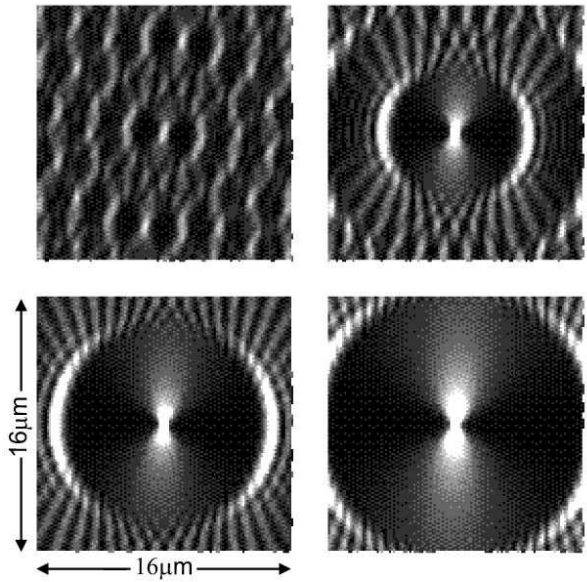}
\caption{\label{fig:funnel2}The optical funnel intensity with lighter regions representing higher intensity.   The rim of the funnel has a larger radius as the number of laser beams is increased.  The panels show 8 beams (top left),16 beams (top right), 24 beams (bottom left), and 32 beams (bottom right). 
 }
\end{figure}

\begin{figure}[b!]
\includegraphics[width=8.5cm]{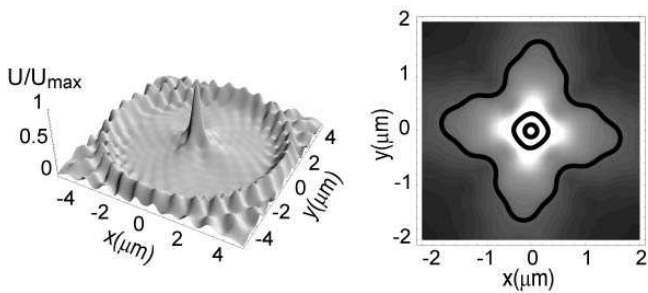}
\caption{\label{fig:funnelpot}The funnel potential(left) and contour plots(right).  Contours for 10$\%$, 50$\%$, and 90$\%$ of the peak intensity are shown.  The optical profile is composed of two noninteracting optical funnels (with different detunings) each composed of 16 laser beams.}
\end{figure}

In this section we discuss an alternative profile that uses traveling waves to avoid the ring structure that mars the applicability of the Bessel profile. We call this structure an optical funnel. 
The central spot of the Bessel squared profile has a FWHM diameter  $w_{\rm FWHM}=0.359 \lambda$.  This small width may be relaxed in exchange for a profile that has a less troublesome ring structure.  One possible optical profile  is the funnel shown in Fig.  \ref{fig:funnel}.  The funnel is a traveling wave field that subtends an angular range of $\pi$ and has amplitudes that decrease linearly on either side of the maximum.  The radius of the rim of the funnel depends  on how many beams are used to create the profile, as shown in Fig.  \ref{fig:funnel2}. All atoms that enter the profile inside  the rim will be funneled towards the point of lowest energy at the center.  The result is a high percentage of the atomic flux being directed  into the central spot. The width of the atomic spot that is written then scales as $w_{a,\rm FWHM}\sim w_{\rm FWHM}\sqrt{ k_B T_a/U_0}$, where  $U_0$
is the maximum well depth of the funnel, and $T_a$ is the temperature of the transverse atomic motion.
  The FWHM of the funnel intensity profile is $0.48 \lambda$ in $x$ and $1.38 \lambda$ in $y$. 
We can create an approximately circular potential by 
  combining two noninteracting funnels (they have a relative detuning that is large compared to $\gamma$, yet small compared to $\Delta$), to get $w_{\rm  FWHM}=0.72 \lambda$ in both the $x$ and $y$ directions, as shown in Fig. \ref{fig:funnelpot}.

\begin{figure}[t!]
\includegraphics[width=8.5cm]{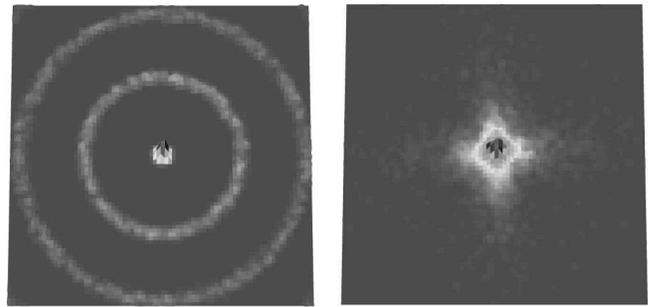}
\caption{\label{fig:besfun}Numerical results for atomic deposition in a Bessel profile (left) and funnel profile (right). Each panel shows a region of $2 ~\mu{\rm m} \times 2~\mu{\rm m}$ and the optical wavelength is 852 nm.
 The atomic beam for both figures had a axial velocity of $14~\rm m/s$, a transverse distribution with $w_{a,\rm FWHM}=1.6~\mu\rm m$, and a transverse temperature of $T_a=22.5~\mu\rm K.$ }
\end{figure}

Even though the width of the symmetrized funnel profile is approximately twice that of the Bessel profile, this optical potential is preferred since it is possible to focus a large percentage of the atomic flux in a single spot with no rings.  Figure \ref{fig:besfun} shows focusing to a single spot using the same atomic beam parameters as in Fig. \ref{fig:bessel}.  For the simulation two noninteracting funnel profiles are placed on top of each other to create a symmetric potential, as described above. The atomic spot in the funnel  has $w_{a, \rm FWHM}=110~{\rm nm}$. Note that the ring structure which is very pronounced when using a Bessel beam has been essentially eliminated.   Both optical profiles have depths of approximately $21 ~\rm mK$. For both profiles cooling beams were used together with the optical focusing as discussed in Sec. \ref{numerics}. 
The results of the simulation show that the funnel captures $94 \%$ of the atoms into the $2~\mu {\rm m} \times 2~\mu {\rm m}$ square, the size of the figure, compared to only $45.3 \%$ for the Bessel profile with only 6\% captured in the central spot.

It is interesting to compare the localized potential created with the funnel to simply using tightly focused Gaussian beams. The use of Gaussian beams would completely eliminate the background ring problem, but comes at the expense of larger spot size. Experiments with a high numerical aperture lens system\cite{Schlos2001} have demonstrated focusing of a single Gaussian beam to a spot diameter of $w_{\rm FWHM}=0.86\lambda.$ A Gaussian beam is highly elongated so we can superpose  two incoherent Gaussians to create a symmetric optical potential.
Because there is a large degree of elongation, even for such a tightly focused Gaussian, the resulting symmetrized intensity profile has $w_{\rm FWHM}=1.4\lambda$ which is almost twice as big as we obtain for the funnel. Comparing Bessel beams, the optical funnel, and Gaussian beams, we see that the funnel combines  a relatively small spot size with large radius rings.
In Sec. \ref{numerics} we demonstrate that two dimensional structures can be written by translating the funnel profile.

%%%%%%%%%%%%%%%%%%%%%%%%%%%%%%%%%%
\section{Atomic source and magnetic precollimation}
\label{MOT}

Most experimental demonstrations of  atomic lithography have used  an oven as a source of thermal atoms. 
The atomic beam is then collimated using mechanical apertures and/or transverse laser cooling to create a beam suitable for lithography experiments. At least two experiments \cite{Fujita1996,Engels1999} have also used
cold or axially cooled atom sources for lithography experiments. A  detailed discussion of the relative merits and requirements of different types of atomic sources for lithography experiments can be found in Ref. \onlinecite{Mesche2003}.   Generally speaking oven based sources provide higher flux and therefore faster writing speeds than cold atom sources. One advantage of cold sources is the low longitudinal velocity which minimizes  surface damage and sputtering  from atom impact on the  deposition substrate. We are interested here in a technique that is suitable for writing feature sizes as small as 100 nm. The requirement for high flux, and fast writing over a large area may therefore be less important than minimization of surface effects. 

The above considerations motivate us to consider the suitability of a cold atom source using the geometry shown schematically in Fig. \ref{fig.schematic}. In order to match a MOT source with a transverse size of $0.1 - 1.0 ~\rm mm$ to the micron sized funnel profile so that secondary rings are eliminated it is necessary to precollimate the atomic beam. We note that this would also be necessary with an oven based source. We propose to do so using a magnetic guide and transverse laser cooling.

The potential $U(x,y)$ due to a  magnetic field $B(x,y)$ in a quadrupole magnetic wave 
guide is

\begin{eqnarray}
\label{magpot}
U(x,y)&=&\mu_B g m  B(x,y)\\
B(x,y)&=&b^{\prime} \sqrt{x^2+y^2}
\end{eqnarray}
where $b^{\prime}=2 \mu_0 J/(\pi a^2),$
 $m$ is the magnetic quantum number, $g$ is the Land\'e factor, $\mu_B$ is the Bohr magneton, $\mu_0$ is the magnetic permeability, $J$ is the current, and $a$ is the distance from the center of the guide to each wire.    As indicated in Fig. \ref{fig.schematic} two-dimensional molasses beams are used to cool the atoms in the magnetic guide.  The resulting size of the atomic beam after cooling is given  by the virial theorem to be 
\begin{equation}
\label{magneticwaveguide}
<r>=\frac{k_B T_a}{g m \mu_B b^{\prime}}.
\end{equation}
Using $T_a=22.5~ \mu \rm K$,  $J=500~\rm A$ and  $a=4~\rm mm$, the average radial distance of the atoms from the center of the guide is $1.34~\mu \rm m$.

The magnetic potential is attractive for atoms that are prepared in states with positive $m.$ 
Since the atoms pass close to the trap axis, we must add a bias field  along the longitudinal axis of the guide  to minimize Majorana spin flips.  The result of this bias field will be a slightly larger atomic beam. The simulations in the next section were done with a bias field $B_0$ to give a total  magnetic field $B(x,y)=\sqrt{B_0^2+b^{\prime^2}(x^2+y^2)}$.  Simulations show that adding a bias field of $B_0=0.1~\rm  G$ in the axial direction  results in a slightly larger  average radial distance of $1.42~ \mu \rm m$ for the above parameters.

%%%%%%%%%%%%%%%%%%%%%%%%%%%%%%%%%%%%%%%%5
\section{Numerical Results}
\label{numerics}

\begin{figure}[!t]
\includegraphics[width=8cm]{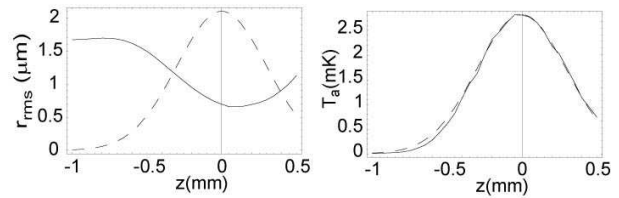}
\caption{\label{fig:funtemp} The RMS width of the atomic beam (left) and transverse temperature (right) of the atoms as they travel through the optical funnel. The initial temperature before entering the funnel is 22.5 $\mu {\rm K}$.  The dashed line shows the axial profile of the funnel potential.}
\end{figure}

Atomic focusing was simulated  using a 4th order Runge-Kutta code to trace the trajectories of individual atoms from the injection into the magnetic wave guide through the optical guide.  The atomic parameters were chosen to correspond to the Cs D2 line ($6^2S_{1/2}-6^2P_{3/2}$) transition with decay rate $\gamma=2\pi\,\times 5.22~\rm MHz$.  Parameter values consistent with magnetic waveguide experiments were chosen \cite{Mandon2000}:  $a=4~\rm mm$ and $J=500~\rm  A$. A small bias field of $B_0=0.1~\rm G$ was included in the simulation as discussed above. 
The atomic beam from the MOT at the entrance to the $20~\rm cm$ long magnetic guide was taken to be a Gaussian distribution with 
$1/e^2$ radius of $100~\mu\rm m$, transverse temperature of $20~\mu\rm  K,$ and  mean longitudinal velocity 
of $14~\rm  m/s.$  Since compression in the magnetic guide heats the atoms we added two dimensional molasses beams to the simulation.  The intensity of the cooling beams was set to $2.9~\rm W/m^2$, or $I/Is=.26$, and $\Delta_m=-\gamma/2$.  The cooling was simulated by randomly changing the momenta of each atom by either $2 \hbar k_m/m_a$,$-2 \hbar k_m/m_a$, or 0 once per scattering time with probabilities of $25\%$, $25\%$, and $50\%$ respectively while constantly damping the atoms with a force\cite{metcalfbook} $-\beta \rm v$ .  Here, $k_m$ is the wavenumber of the molasses beams, $m_a$ is the mass of the atom, and $\beta$ is the damping coefficient. 
The velocity kick was added in both transverse directions to independently cool along each axis.  The linewidth $\gamma$, was decreased by a factor of 8 and the recoil velocity was decreased to 0.23 cm/s to simulate sub-Doppler cooling to approximately $22.5~\mu\rm  K$ in the transverse plane. The resulting FWHM of the atomic beam after cooling but still in the magnetic wave guide is $w_{a,\rm FWHM}\simeq 1.68 ~\mu\rm  m$.  

At the end of the magnetic wave guide we feed the atoms into the optical funnel. The funnel was comprised of two noninteracting funnel profiles, each consisting of 32 laser beams.  The two funnels are overlapped to form a  symmetric potential, as in Fig \ref{fig:funnelpot}.  The axial profile of the beams was 
  $g(z)=\exp(-2 z^2/w_z^2)$, with  $w_z=0.6~\rm  mm$ and their cross section was  assumed to be circular in the focusing region.  A peak intensity of $\rm7 \times 10^6~\rm W/m^2$ and $\Delta/2\pi = - 10 ~ \rm GHz$ was chosen for one funnel and $\rm6.3 \times 10^6~\rm W/m^2$ and $\Delta/2\pi = - 9 ~ \rm GHz$ for the other.  This choice of parameters gives the same well depth for both funnels, but detuned such that they do not interact.  This results in a total well depth of $U_0/k_B=21~\rm  mK$ with a laser power requirement of approximately 124  and 111 mW respectively.

\begin{figure}[!t]
\includegraphics[width=7cm]{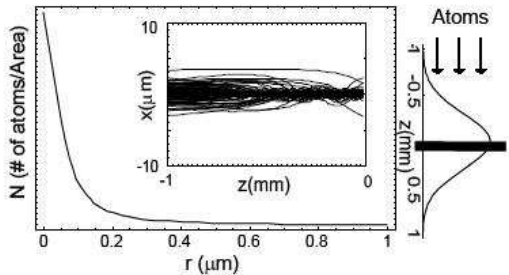}
\caption{\label{fig:traj} Number density of the optically focused atomic beam with $w_{a,\rm FWHM}= 110 ~ \rm nm.$  The insert shows the trajectories of 100 randomly chosen  atoms as they travel through the optical profile.  The compression of the beam is clearly seen. The substrate, shown on the right, is in the center of the Gaussian profile.}
\end{figure}

\begin{figure}[b!]
\includegraphics[width=7cm]{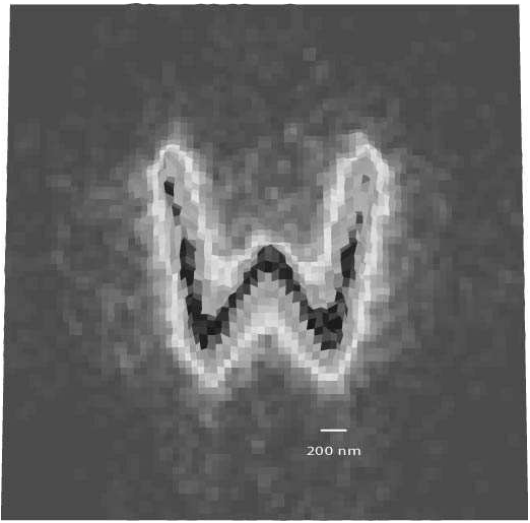}
\caption{\label{fig:W}Distribution of atoms deposited to write the  letter  W  written by changing the phases of the beams creating the  optical profile.  The magnetic guide is stationary and aligned with the center of the picture which shows a region of size $4 ~\mu{\rm m} \times 4~\mu \rm m $.}
\end{figure}

Since the atoms are heated as they travel into the  optical funnel as shown in Fig.  \ref{fig:funtemp}, we add cooling beams to the optical profile.   For the cooling beams, the detuning was set to half the line width and the intensity of each beam was 40~\% of the saturation intensity.  Two pairs of orthogonal cooling beams  were used with the same axial spatial mode as the beams used to create the funnel.  The cooling beams help to limit the temperature, but do not interact strongly enough with the atoms to cool them to the Doppler temperature before the atoms hit the substrate. The substrate is placed at the center of the Gaussian profile as has been done in some experiments\cite{Lison1997}.  The transverse temperature of the atoms just before the substrate is approximately 2.5 mK.   Since the atoms are heated to well above the Doppler temperature when entering the optical profile, cooling was simulated solely by a damping force, ignoring photon kicks.  The resulting number density in the optical profile has $w_{a,\rm FWHM}= 110~\rm  nm$ with 82\% of the atoms falling within a radius of 500 nm and 20\% falling within the FWHM of the beam.  The trajectories of 100 atoms are shown in the inset of Fig. \ref{fig:traj}.

Due to the large intensity of the funnel beams additional heating due to photon scattering  may also be of concern.  The peak scattering rate  at the center of the many beam optical funnel is \cite{metcalfbook}
\begin{equation}
\label{scatter}
r\sim \frac{\gamma}{2}\frac{\frac{I_0}{I_s}}{1+\frac{4\Delta^2}{\gamma^2}+\frac{I_0}{I_s}}
\end{equation}
For the funnel parameters given above this results in a maximum scattering 
rate of $r=1.3 \times 10^6~\rm s^{-1}.$  The atom-light interaction time is approximately 0.1 ms and results in  scattering of about  150 photons. The scattering of these photons adds  $\frac{m_a}{k_B}(\frac{\sqrt{150} \hbar k}{m_a})^2=30~\mu \rm K$ to the atomic temperature which  can be neglected since the temperature of the atoms at the center of the profile is approximately $2.5~\rm  mK.$

\begin{figure}[t!]
\includegraphics[width=7cm]{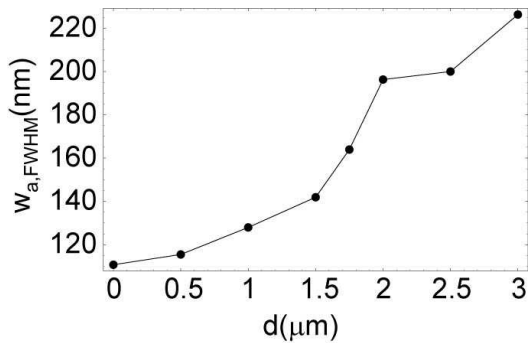}
\caption{\label{fig:fwhm} FWHM of the atomic beam at the substrate as a function of the distance $d$ between the magnetic waveguide and the funnel.}
\end{figure}

 To simulate writing of a two-dimensional pattern the optical funnel is  reconstructed at varying distances from the atomic beam axis using the phases given by Eq. (\ref{phase}).  A simulation of a W is shown in Fig. \ref{fig:W}.  The W is made by positioning the funnel at 101 different spots and depositing a total of 24240 atoms.  The FWHM of the atomic beam at the substrate increases the further the optical funnel  is from the center of the magnetic guide.  This results in a small spreading of the ends of the W, visible in Fig.  \ref{fig:W}. Figure  \ref{fig:fwhm}  shows that the FWHM  of the atomic beam on the substrate increases by about a factor of two when the funnel is moved by a distance of $3~\mu\rm m.$   This  is because the atoms that enter the potential near the rim obtain a larger radial velocity while traveling to the potential minimum.  As a result, the atoms are hotter when they arrive at the substrate and thus have a larger $w_{a,\rm FWHM}. $

In order for this type of lithography to be practical the writing speed must not be too slow. 
  Coverage of a surface with one monolayer of Cs corresponds to a surface density of about  \cite{Lison1997}  $4 \times 10^{15}~\rm  atoms /cm^2$. It has been shown\cite{Berggr1997} that between 3 and 7 monolayers of Cs are needed to create enough damage for exposure of organic   self-assembled monolayer (SAM) coatings.
  Given a spot of $w_{a,\rm FWHM}=  110~\rm  nm$, which defines a pixel with area $\pi 55^2=9500~\rm nm^2$, the exposure per pixel for one monolayer of Cs is $3.8 \times 10^5$ atoms/pixel.  Given a flux of $5\times 10^8
~\rm  atoms/s$  and 20\% of the atoms falling within an area of diameter $w_{a,\rm FWHM}$, it will take about $3.8~\rm ms$  to deposit one monolayer of Cs.  The time needed to successfully write 3 to 7 monolayers of cesium to a pixel is then between $11$ and $27~\rm ms$.  The W has lengths which total $4.8~\mu \rm m$ which corresponds to at least $4.8~\mu {\rm m}/110~\rm nm \approx 44~\rm spots$.  Therefore, the writing time for the W would be approximately 1 second.

\section{Discussion}
\label{sec7}

We have  described an atomic lithography system that uses magnetic and optical fields 
to focus  atoms from a MOT onto a sub-micron spot. Doing so requires combining  confining potentials with optical molasses to maintain low atomic temperatures.   Using numerical simulations of experimentally realistic parameters we produce pixels with  $w_{a,\rm FWHM}=110~\rm nm$, and a writing speed of order $20~\rm ms$ per pixel.  The position of the pixel  can be scanned to write arbitrary planar structures using phase shifts created by a  SLM. The optical funnel that produces the final focusing has an acceptance region with diameter of about  $8~\rm \mu\rm m.$ However, the funnel cannot be scanned that far since when the funnel position moves a distance comparable to the width of the beam leaving the magnetic guide the spot size created by the funnel starts to increase. We can therefore say that for the parameters investigated it appears possible to write no more than several hundred independent pixels with $w_{a,\rm FWHM}\sim 110~\rm nm$. Alternatively the system could be optimized to write a single  small, but stationary spot, and the substrate scanned mechanically.
This would in principle allow an arbitrarily large number of pixels to be written.

A long term goal of atomic lithography is to produce a large scale lithographic process. 
To do so a number of challenges will have to be overcome. 
The optical funnel will only capture atoms into a small spot if they enter the funnel close enough to the center.  Theoretically, we could  increase the axial thickness of the funnel,  or increase the laser intensity to increase the range at which the funnel will capture atoms.  Experimentally, this is limited by available laser power.   Another experimental difficulty will be the sub-Doppler cooling that is required in the magnetic wave guide.  Others have shown theoretically that sub-Doppler cooling is possible with small 
spot sizes\cite{Balyki2003}, but experiments have not yet demonstrated  this level of cooling of a traveling beam in a quadrupole wave guide.  Requirements on cooling  efficiency could be traded off against stronger compression due to larger magnetic fields. This is limited by the ability to run large currents through small wires. Approaches based on
lithographically patterned wires\cite{Hinds1999} which will enable waveguides with smaller dimensions may enable even tighter confinement. 

In conclusion we expect that solutions to these issues, as well as further optimization of performance by refinement of parameters will be possible. Full evaluation of the suitability of the atom-optical approach  for writing complex structures described here will ultimately rely on experimental tests.

The authors thank Deniz Yavuz for helpful  discussions.
This work was supported by NSF grant PHY-0210357,
and an Advanced Opportunity Fellowship from the University of Wisconsin graduate school.

\bibliography{atomiclithography}

\begin{thebibliography}{32}
\expandafter\ifx\csname natexlab\endcsname\relax\def\natexlab#1{#1}\fi
\expandafter\ifx\csname bibnamefont\endcsname\relax
  \def\bibnamefont#1{#1}\fi
\expandafter\ifx\csname bibfnamefont\endcsname\relax
  \def\bibfnamefont#1{#1}\fi
\expandafter\ifx\csname citenamefont\endcsname\relax
  \def\citenamefont#1{#1}\fi
\expandafter\ifx\csname url\endcsname\relax
  \def\url#1{\texttt{#1}}\fi
\expandafter\ifx\csname urlprefix\endcsname\relax\def\urlprefix{URL }\fi
\providecommand{\bibinfo}[2]{#2}
\providecommand{\eprint}[2][]{\url{#2}}

\bibitem[{\citenamefont{Timp et~al.}(1992)\citenamefont{Timp, Behringer,
  Tennant, Cunningham, Prentiss, and Berggren}}]{TIMP1992}
\bibinfo{author}{\bibfnamefont{G.}~\bibnamefont{Timp}},
  \bibinfo{author}{\bibfnamefont{R.~E.} \bibnamefont{Behringer}},
  \bibinfo{author}{\bibfnamefont{D.~M.} \bibnamefont{Tennant}},
  \bibinfo{author}{\bibfnamefont{J.~E.} \bibnamefont{Cunningham}},
  \bibinfo{author}{\bibfnamefont{M.}~\bibnamefont{Prentiss}}, \bibnamefont{and}
  \bibinfo{author}{\bibfnamefont{K.~K.} \bibnamefont{Berggren}},
  \bibinfo{journal}{Phys. Rev. Lett.} \textbf{\bibinfo{volume}{69}},
  \bibinfo{pages}{1636} (\bibinfo{year}{1992}).

\bibitem[{\citenamefont{McClelland et~al.}(1993)\citenamefont{McClelland,
  Scholten, Palm, and Celotta}}]{MCCLEL1993}
\bibinfo{author}{\bibfnamefont{J.~J.} \bibnamefont{McClelland}},
  \bibinfo{author}{\bibfnamefont{R.~E.} \bibnamefont{Scholten}},
  \bibinfo{author}{\bibfnamefont{E.~C.} \bibnamefont{Palm}}, \bibnamefont{and}
  \bibinfo{author}{\bibfnamefont{R.~J.} \bibnamefont{Celotta}},
  \bibinfo{journal}{Science} \textbf{\bibinfo{volume}{262}},
  \bibinfo{pages}{877} (\bibinfo{year}{1993}).

\bibitem[{\citenamefont{Schulze et~al.}(2000)\citenamefont{Schulze, Brezger,
  Mertens, Pivk, Pfau, and Mlynek}}]{Schulz2000}
\bibinfo{author}{\bibfnamefont{T.}~\bibnamefont{Schulze}},
  \bibinfo{author}{\bibfnamefont{B.}~\bibnamefont{Brezger}},
  \bibinfo{author}{\bibfnamefont{R.}~\bibnamefont{Mertens}},
  \bibinfo{author}{\bibfnamefont{M.}~\bibnamefont{Pivk}},
  \bibinfo{author}{\bibfnamefont{T.}~\bibnamefont{Pfau}}, \bibnamefont{and}
  \bibinfo{author}{\bibfnamefont{J.}~\bibnamefont{Mlynek}},
  \bibinfo{journal}{Appl. Phys. B} \textbf{\bibinfo{volume}{70}},
  \bibinfo{pages}{671} (\bibinfo{year}{2000}).

\bibitem[{\citenamefont{Bradley et~al.}(1999)\citenamefont{Bradley, Anderson,
  McClelland, and Celotta}}]{Bradle1999}
\bibinfo{author}{\bibfnamefont{C.~C.} \bibnamefont{Bradley}},
  \bibinfo{author}{\bibfnamefont{W.~R.} \bibnamefont{Anderson}},
  \bibinfo{author}{\bibfnamefont{J.~J.} \bibnamefont{McClelland}},
  \bibnamefont{and} \bibinfo{author}{\bibfnamefont{R.~J.}
  \bibnamefont{Celotta}}, \bibinfo{journal}{Appl. Surf. Sci.}
  \textbf{\bibinfo{volume}{141}}, \bibinfo{pages}{210} (\bibinfo{year}{1999}).

\bibitem[{\citenamefont{Petra et~al.}(2004)\citenamefont{Petra, van Leeuwen,
  Feenstra, Hogervorst, and Vassen}}]{Petra2004}
\bibinfo{author}{\bibfnamefont{S.~J.~H.} \bibnamefont{Petra}},
  \bibinfo{author}{\bibfnamefont{K.~A.~H.} \bibnamefont{van Leeuwen}},
  \bibinfo{author}{\bibfnamefont{L.}~\bibnamefont{Feenstra}},
  \bibinfo{author}{\bibfnamefont{W.}~\bibnamefont{Hogervorst}},
  \bibnamefont{and} \bibinfo{author}{\bibfnamefont{W.}~\bibnamefont{Vassen}},
  \bibinfo{journal}{Appl. Phys. B} \textbf{\bibinfo{volume}{79}},
  \bibinfo{pages}{279} (\bibinfo{year}{2004}).

\bibitem[{\citenamefont{Kaenders et~al.}(1995)\citenamefont{Kaenders, Lison,
  Richter, Wynands, and Meschede}}]{Kaenders1995}
\bibinfo{author}{\bibfnamefont{W.~G.} \bibnamefont{Kaenders}},
  \bibinfo{author}{\bibfnamefont{F.}~\bibnamefont{Lison}},
  \bibinfo{author}{\bibfnamefont{A.}~\bibnamefont{Richter}},
  \bibinfo{author}{\bibfnamefont{R.}~\bibnamefont{Wynands}}, \bibnamefont{and}
  \bibinfo{author}{\bibfnamefont{D.}~\bibnamefont{Meschede}},
  \bibinfo{journal}{Nature} \textbf{\bibinfo{volume}{375}},
  \bibinfo{pages}{214} (\bibinfo{year}{1995}).

\bibitem[{\citenamefont{Hinds and Hughes}(1999)}]{Hinds1999}
\bibinfo{author}{\bibfnamefont{E.~A.} \bibnamefont{Hinds}} \bibnamefont{and}
  \bibinfo{author}{\bibfnamefont{I.~G.} \bibnamefont{Hughes}},
  \bibinfo{journal}{J. Phys. D: Appl. Phys.} \textbf{\bibinfo{volume}{32}},
  \bibinfo{pages}{R119} (\bibinfo{year}{1999}).

\bibitem[{\citenamefont{Oberthaler and Pfau}(2003)}]{Oberth2003}
\bibinfo{author}{\bibfnamefont{M.~K.} \bibnamefont{Oberthaler}}
  \bibnamefont{and} \bibinfo{author}{\bibfnamefont{T.}~\bibnamefont{Pfau}},
  \bibinfo{journal}{J. Phys. Condens. Matter} \textbf{\bibinfo{volume}{15}},
  \bibinfo{pages}{R233} (\bibinfo{year}{2003}).

\bibitem[{\citenamefont{Meschede and Metcalf}(2003)}]{Mesche2003}
\bibinfo{author}{\bibfnamefont{D.}~\bibnamefont{Meschede}} \bibnamefont{and}
  \bibinfo{author}{\bibfnamefont{H.}~\bibnamefont{Metcalf}},
  \bibinfo{journal}{J Phys D: Appl Phys} \textbf{\bibinfo{volume}{36}},
  \bibinfo{pages}{R17} (\bibinfo{year}{2003}).

\bibitem[{\citenamefont{McClelland et~al.}(2004)\citenamefont{McClelland, Hill,
  Pichler, and Celotta}}]{McClel2004}
\bibinfo{author}{\bibfnamefont{J.~J.} \bibnamefont{McClelland}},
  \bibinfo{author}{\bibfnamefont{S.~B.} \bibnamefont{Hill}},
  \bibinfo{author}{\bibfnamefont{M.}~\bibnamefont{Pichler}}, \bibnamefont{and}
  \bibinfo{author}{\bibfnamefont{R.~J.} \bibnamefont{Celotta}},
  \bibinfo{journal}{Science and Technology of Adv. Mater.}
  \textbf{\bibinfo{volume}{5}}, \bibinfo{pages}{575} (\bibinfo{year}{2004}).

\bibitem[{\citenamefont{Jurdik et~al.}(2004)\citenamefont{Jurdik, Myszkiewicz,
  Hohlfeld, Tsukamoto, Toonen, van Etteger, Gerritsen, Hermsen,
  Goldbach-Aschemann, Meerts et~al.}}]{Jurdik2004}
\bibinfo{author}{\bibfnamefont{E.}~\bibnamefont{Jurdik}},
  \bibinfo{author}{\bibfnamefont{G.}~\bibnamefont{Myszkiewicz}},
  \bibinfo{author}{\bibfnamefont{J.}~\bibnamefont{Hohlfeld}},
  \bibinfo{author}{\bibfnamefont{A.}~\bibnamefont{Tsukamoto}},
  \bibinfo{author}{\bibfnamefont{A.~J.} \bibnamefont{Toonen}},
  \bibinfo{author}{\bibfnamefont{A.~F.} \bibnamefont{van Etteger}},
  \bibinfo{author}{\bibfnamefont{J.}~\bibnamefont{Gerritsen}},
  \bibinfo{author}{\bibfnamefont{J.}~\bibnamefont{Hermsen}},
  \bibinfo{author}{\bibfnamefont{S.}~\bibnamefont{Goldbach-Aschemann}},
  \bibinfo{author}{\bibfnamefont{W.~L.} \bibnamefont{Meerts}},
  \bibnamefont{et~al.}, \bibinfo{journal}{Phys. Rev. B}
  \textbf{\bibinfo{volume}{69}}, \bibinfo{pages}{201102(R)}
  (\bibinfo{year}{2004}).

\bibitem[{\citenamefont{M{\"u}tzel et~al.}(2002)\citenamefont{M{\"u}tzel,
  Tandler, Haubrich, Meschede, Peithmann, Flasp{\"o}hler, and
  Buse}}]{Mutzel2002}
\bibinfo{author}{\bibfnamefont{M.}~\bibnamefont{M{\"u}tzel}},
  \bibinfo{author}{\bibfnamefont{S.}~\bibnamefont{Tandler}},
  \bibinfo{author}{\bibfnamefont{D.}~\bibnamefont{Haubrich}},
  \bibinfo{author}{\bibfnamefont{D.}~\bibnamefont{Meschede}},
  \bibinfo{author}{\bibfnamefont{K.}~\bibnamefont{Peithmann}},
  \bibinfo{author}{\bibfnamefont{M.}~\bibnamefont{Flasp{\"o}hler}},
  \bibnamefont{and} \bibinfo{author}{\bibfnamefont{K.}~\bibnamefont{Buse}},
  \bibinfo{journal}{Phys. Rev. Lett.} \textbf{\bibinfo{volume}{88}},
  \bibinfo{pages}{083601} (\bibinfo{year}{2002}).

\bibitem[{\citenamefont{M{\"u}tzel et~al.}(2003)\citenamefont{M{\"u}tzel,
  Rasbach, Meschede, Burstedde, Braun, Kunoth, Peithmann, and
  Buse}}]{Mutzel2003}
\bibinfo{author}{\bibfnamefont{M.}~\bibnamefont{M{\"u}tzel}},
  \bibinfo{author}{\bibfnamefont{U.}~\bibnamefont{Rasbach}},
  \bibinfo{author}{\bibfnamefont{D.}~\bibnamefont{Meschede}},
  \bibinfo{author}{\bibfnamefont{C.}~\bibnamefont{Burstedde}},
  \bibinfo{author}{\bibfnamefont{J.}~\bibnamefont{Braun}},
  \bibinfo{author}{\bibfnamefont{A.}~\bibnamefont{Kunoth}},
  \bibinfo{author}{\bibfnamefont{K.}~\bibnamefont{Peithmann}},
  \bibnamefont{and} \bibinfo{author}{\bibfnamefont{K.}~\bibnamefont{Buse}},
  \bibinfo{journal}{Appl. Phys. B} \textbf{\bibinfo{volume}{77}},
  \bibinfo{pages}{1} (\bibinfo{year}{2003}).

\bibitem[{\citenamefont{Oberst et~al.}(2003)\citenamefont{Oberst, Kasashima,
  Balykin, and Shimizu}}]{Oberst2003}
\bibinfo{author}{\bibfnamefont{H.}~\bibnamefont{Oberst}},
  \bibinfo{author}{\bibfnamefont{S.}~\bibnamefont{Kasashima}},
  \bibinfo{author}{\bibfnamefont{V.~I.} \bibnamefont{Balykin}},
  \bibnamefont{and} \bibinfo{author}{\bibfnamefont{F.}~\bibnamefont{Shimizu}},
  \bibinfo{journal}{Phys. Rev. A} \textbf{\bibinfo{volume}{68}},
  \bibinfo{pages}{013606} (\bibinfo{year}{2003}).

\bibitem[{\citenamefont{Renn et~al.}(1995)\citenamefont{Renn, Montgomery,
  Vdovin, Anderson, Wieman, and Cornell}}]{Renn1995}
\bibinfo{author}{\bibfnamefont{M.~J.} \bibnamefont{Renn}},
  \bibinfo{author}{\bibfnamefont{D.}~\bibnamefont{Montgomery}},
  \bibinfo{author}{\bibfnamefont{O.}~\bibnamefont{Vdovin}},
  \bibinfo{author}{\bibfnamefont{D.~Z.} \bibnamefont{Anderson}},
  \bibinfo{author}{\bibfnamefont{C.~E.} \bibnamefont{Wieman}},
  \bibnamefont{and} \bibinfo{author}{\bibfnamefont{E.~A.}
  \bibnamefont{Cornell}}, \bibinfo{journal}{Phys. Rev. Lett.}
  \textbf{\bibinfo{volume}{75}}, \bibinfo{pages}{3253} (\bibinfo{year}{1995}).

\bibitem[{\citenamefont{Bjorkholm et~al.}(1978)\citenamefont{Bjorkholm,
  Freeman, Ashkin, and Pearson}}]{Bjorkh1978}
\bibinfo{author}{\bibfnamefont{J.~E.} \bibnamefont{Bjorkholm}},
  \bibinfo{author}{\bibfnamefont{R.~R.} \bibnamefont{Freeman}},
  \bibinfo{author}{\bibfnamefont{A.}~\bibnamefont{Ashkin}}, \bibnamefont{and}
  \bibinfo{author}{\bibfnamefont{D.~B.} \bibnamefont{Pearson}},
  \bibinfo{journal}{Phys. Rev. Lett.} \textbf{\bibinfo{volume}{41}},
  \bibinfo{pages}{1361} (\bibinfo{year}{1978}).

\bibitem[{\citenamefont{Balykin and Letokhov}(1987)}]{Balyki1987}
\bibinfo{author}{\bibfnamefont{V.~I.} \bibnamefont{Balykin}} \bibnamefont{and}
  \bibinfo{author}{\bibfnamefont{V.~S.} \bibnamefont{Letokhov}},
  \bibinfo{journal}{Opt. Commun.} \textbf{\bibinfo{volume}{64}},
  \bibinfo{pages}{151} (\bibinfo{year}{1987}).

\bibitem[{\citenamefont{Dubetsky and Berman}(1998)}]{Dubets1998}
\bibinfo{author}{\bibfnamefont{B.}~\bibnamefont{Dubetsky}} \bibnamefont{and}
  \bibinfo{author}{\bibfnamefont{P.~R.} \bibnamefont{Berman}},
  \bibinfo{journal}{Phys. Rev. A} \textbf{\bibinfo{volume}{58}},
  \bibinfo{pages}{2413} (\bibinfo{year}{1998}).

\bibitem[{\citenamefont{Okamoto et~al.}(2001)\citenamefont{Okamoto, Inouye, and
  Kawata}}]{Okamo2001}
\bibinfo{author}{\bibfnamefont{K.}~\bibnamefont{Okamoto}},
  \bibinfo{author}{\bibfnamefont{Y.}~\bibnamefont{Inouye}}, \bibnamefont{and}
  \bibinfo{author}{\bibfnamefont{S.}~\bibnamefont{Kawata}},
  \bibinfo{journal}{Jpn. J. Appl. Phys. Part 1} \textbf{\bibinfo{volume}{40}},
  \bibinfo{pages}{4544} (\bibinfo{year}{2001}).

\bibitem[{\citenamefont{Lu et~al.}(1996)\citenamefont{Lu, Corwin, Renn,
  Anderson, Cornell, and Wieman}}]{Lu1996}
\bibinfo{author}{\bibfnamefont{Z.~T.} \bibnamefont{Lu}},
  \bibinfo{author}{\bibfnamefont{K.~L.} \bibnamefont{Corwin}},
  \bibinfo{author}{\bibfnamefont{M.~J.} \bibnamefont{Renn}},
  \bibinfo{author}{\bibfnamefont{M.~H.} \bibnamefont{Anderson}},
  \bibinfo{author}{\bibfnamefont{E.~A.} \bibnamefont{Cornell}},
  \bibnamefont{and} \bibinfo{author}{\bibfnamefont{C.~E.}
  \bibnamefont{Wieman}}, \bibinfo{journal}{Phys. Rev. Lett.}
  \textbf{\bibinfo{volume}{77}}, \bibinfo{pages}{3331} (\bibinfo{year}{1996}).

\bibitem[{\citenamefont{Mandonnet et~al.}(2000)\citenamefont{Mandonnet,
  Minguzzi, Dum, Carusotto, Castin, and Dalibard}}]{Mandon2000}
\bibinfo{author}{\bibfnamefont{E.}~\bibnamefont{Mandonnet}},
  \bibinfo{author}{\bibfnamefont{A.}~\bibnamefont{Minguzzi}},
  \bibinfo{author}{\bibfnamefont{R.}~\bibnamefont{Dum}},
  \bibinfo{author}{\bibfnamefont{I.}~\bibnamefont{Carusotto}},
  \bibinfo{author}{\bibfnamefont{Y.}~\bibnamefont{Castin}}, \bibnamefont{and}
  \bibinfo{author}{\bibfnamefont{J.}~\bibnamefont{Dalibard}},
  \bibinfo{journal}{Eur. Phys. J. D} \textbf{\bibinfo{volume}{10}},
  \bibinfo{pages}{9} (\bibinfo{year}{2000}).

\bibitem[{\citenamefont{McClelland}(1995)}]{McClel1995}
\bibinfo{author}{\bibfnamefont{J.~J.} \bibnamefont{McClelland}},
  \bibinfo{journal}{J. Opt. Soc. Am. B} \textbf{\bibinfo{volume}{12}},
  \bibinfo{pages}{1761} (\bibinfo{year}{1995}).

\bibitem[{\citenamefont{McGloin et~al.}(2003)\citenamefont{McGloin, Spalding,
  Melville, Sibbett, and Dholakia}}]{McGloi2003}
\bibinfo{author}{\bibfnamefont{D.}~\bibnamefont{McGloin}},
  \bibinfo{author}{\bibfnamefont{G.~C.} \bibnamefont{Spalding}},
  \bibinfo{author}{\bibfnamefont{H.}~\bibnamefont{Melville}},
  \bibinfo{author}{\bibfnamefont{W.}~\bibnamefont{Sibbett}}, \bibnamefont{and}
  \bibinfo{author}{\bibfnamefont{K.}~\bibnamefont{Dholakia}},
  \bibinfo{journal}{Opt. Express} \textbf{\bibinfo{volume}{11}},
  \bibinfo{pages}{158} (\bibinfo{year}{2003}).

\bibitem[{\citenamefont{Bergamini et~al.}(2004)\citenamefont{Bergamini,
  Darqui\'e, Jones, Jacubowiez, Browaeys, and Grangier}}]{Bergam2004}
\bibinfo{author}{\bibfnamefont{S.}~\bibnamefont{Bergamini}},
  \bibinfo{author}{\bibfnamefont{B.}~\bibnamefont{Darqui\'e}},
  \bibinfo{author}{\bibfnamefont{M.}~\bibnamefont{Jones}},
  \bibinfo{author}{\bibfnamefont{L.}~\bibnamefont{Jacubowiez}},
  \bibinfo{author}{\bibfnamefont{A.}~\bibnamefont{Browaeys}}, \bibnamefont{and}
  \bibinfo{author}{\bibfnamefont{P.}~\bibnamefont{Grangier}},
  \bibinfo{journal}{J. Opt. Soc. Am. B} \textbf{\bibinfo{volume}{21}},
  \bibinfo{pages}{1889} (\bibinfo{year}{2004}).

\bibitem[{\citenamefont{Saffman}(2004)}]{saffma2004}
\bibinfo{author}{\bibfnamefont{M.}~\bibnamefont{Saffman}},
  \bibinfo{journal}{Opt. Lett.} \textbf{\bibinfo{volume}{29}},
  \bibinfo{pages}{1016} (\bibinfo{year}{2004}).

\bibitem[{\citenamefont{Schlosser et~al.}(2001)\citenamefont{Schlosser,
  Reymond, Protsenko, and Grangier}}]{Schlos2001}
\bibinfo{author}{\bibfnamefont{N.}~\bibnamefont{Schlosser}},
  \bibinfo{author}{\bibfnamefont{G.}~\bibnamefont{Reymond}},
  \bibinfo{author}{\bibfnamefont{I.}~\bibnamefont{Protsenko}},
  \bibnamefont{and} \bibinfo{author}{\bibfnamefont{P.}~\bibnamefont{Grangier}},
  \bibinfo{journal}{Nature} \textbf{\bibinfo{volume}{411}},
  \bibinfo{pages}{1024} (\bibinfo{year}{2001}).

\bibitem[{\citenamefont{Fujita et~al.}(1996)\citenamefont{Fujita, Morinaga,
  Kishimoto, Yasuda, Matsui, and Shimizu}}]{Fujita1996}
\bibinfo{author}{\bibfnamefont{J.}~\bibnamefont{Fujita}},
  \bibinfo{author}{\bibfnamefont{M.}~\bibnamefont{Morinaga}},
  \bibinfo{author}{\bibfnamefont{T.}~\bibnamefont{Kishimoto}},
  \bibinfo{author}{\bibfnamefont{M.}~\bibnamefont{Yasuda}},
  \bibinfo{author}{\bibfnamefont{S.}~\bibnamefont{Matsui}}, \bibnamefont{and}
  \bibinfo{author}{\bibfnamefont{F.}~\bibnamefont{Shimizu}},
  \bibinfo{journal}{Nature} \textbf{\bibinfo{volume}{380}},
  \bibinfo{pages}{691} (\bibinfo{year}{1996}).

\bibitem[{\citenamefont{Engels et~al.}(1999)\citenamefont{Engels, Salewski,
  Levsen, Sengstock, and Ertmer}}]{Engels1999}
\bibinfo{author}{\bibfnamefont{P.}~\bibnamefont{Engels}},
  \bibinfo{author}{\bibfnamefont{S.}~\bibnamefont{Salewski}},
  \bibinfo{author}{\bibfnamefont{H.}~\bibnamefont{Levsen}},
  \bibinfo{author}{\bibfnamefont{K.}~\bibnamefont{Sengstock}},
  \bibnamefont{and} \bibinfo{author}{\bibfnamefont{W.}~\bibnamefont{Ertmer}},
  \bibinfo{journal}{Appl. Phys. B} \textbf{\bibinfo{volume}{69}},
  \bibinfo{pages}{407} (\bibinfo{year}{1999}).

\bibitem[{\citenamefont{Metcalf and van~der Straten}(1999)}]{metcalfbook}
\bibinfo{author}{\bibfnamefont{H.~J.} \bibnamefont{Metcalf}} \bibnamefont{and}
  \bibinfo{author}{\bibfnamefont{P.}~\bibnamefont{van~der Straten}},
  \emph{\bibinfo{title}{Laser Cooling and Trapping}}
  (\bibinfo{publisher}{Springer-Verlag, New York}, \bibinfo{year}{1999}).

\bibitem[{\citenamefont{Lison et~al.}(1997)\citenamefont{Lison, Adams,
  Haubrich, Kreis, Nowak, and Meschede}}]{Lison1997}
\bibinfo{author}{\bibfnamefont{F.}~\bibnamefont{Lison}},
  \bibinfo{author}{\bibfnamefont{H.-J.} \bibnamefont{Adams}},
  \bibinfo{author}{\bibfnamefont{D.}~\bibnamefont{Haubrich}},
  \bibinfo{author}{\bibfnamefont{M.}~\bibnamefont{Kreis}},
  \bibinfo{author}{\bibfnamefont{S.}~\bibnamefont{Nowak}}, \bibnamefont{and}
  \bibinfo{author}{\bibfnamefont{D.}~\bibnamefont{Meschede}},
  \bibinfo{journal}{Appl. Phys. B} \textbf{\bibinfo{volume}{65}},
  \bibinfo{pages}{419} (\bibinfo{year}{1997}).

\bibitem[{\citenamefont{Berggren et~al.}(1997)\citenamefont{Berggren, Younkin,
  Cheung, Prentiss, Black, Whitesides, Ralph, Black, and Tinkham}}]{Berggr1997}
\bibinfo{author}{\bibfnamefont{K.~K.} \bibnamefont{Berggren}},
  \bibinfo{author}{\bibfnamefont{R.}~\bibnamefont{Younkin}},
  \bibinfo{author}{\bibfnamefont{E.}~\bibnamefont{Cheung}},
  \bibinfo{author}{\bibfnamefont{M.}~\bibnamefont{Prentiss}},
  \bibinfo{author}{\bibfnamefont{A.~J.} \bibnamefont{Black}},
  \bibinfo{author}{\bibfnamefont{G.~M.} \bibnamefont{Whitesides}},
  \bibinfo{author}{\bibfnamefont{D.~C.} \bibnamefont{Ralph}},
  \bibinfo{author}{\bibfnamefont{C.~T.} \bibnamefont{Black}}, \bibnamefont{and}
  \bibinfo{author}{\bibfnamefont{M.}~\bibnamefont{Tinkham}},
  \bibinfo{journal}{Adv. Mater.} \textbf{\bibinfo{volume}{9}},
  \bibinfo{pages}{52} (\bibinfo{year}{1997}).

\bibitem[{\citenamefont{Balykin and Minogin}(2003)}]{Balyki2003}
\bibinfo{author}{\bibfnamefont{V.~I.} \bibnamefont{Balykin}} \bibnamefont{and}
  \bibinfo{author}{\bibfnamefont{V.~G.} \bibnamefont{Minogin}},
  \bibinfo{journal}{J. Exp. Theor. Phys.} \textbf{\bibinfo{volume}{96}},
  \bibinfo{pages}{8} (\bibinfo{year}{2003}), \bibinfo{note}{translated from Zh.
  {\'E}ksp. Teor. Fiz. {\bf 123}, 13 (2003)}.

\end{thebibliography}
\end{document}